\documentclass[aps,pre,epsfig,floats,twocolumn,superscriptaddress,amssymb,amsmath,showpacs]{revtex4}
\usepackage{graphicx}
\begin{document}
\title{Orientational Ordering and Dynamics of Rodlike Polyelectrolytes}

\author{Hossein Fazli}

\affiliation{Institute for Advanced Studies in Basic Sciences,
Zanjan 45195-159, Iran}

\author{Ramin Golestanian}

\affiliation{Institute for Advanced Studies in Basic Sciences,
Zanjan 45195-159, Iran}

\author{Mohammad R. Kolahchi}

\affiliation{Institute for Advanced Studies in Basic Sciences,
Zanjan 45195-159, Iran}

\date{\today}

\begin{abstract}
The interplay between electrostatic interactions and orientational
correlations is studied for a model system of charged rods
positioned on a chain, using Monte Carlo simulation techniques. It
is shown that the coupling brings about the notion of {\em
electrostatic frustration}, which in turn results in: (i) a rich
variety of novel orientational orderings such as chiral phases,
and (ii) an inherently slow dynamics characterized by
stretched-exponential behavior in the relaxation functions of the
system.

\end{abstract}

 \maketitle

\section{Introduction}  \label{sec:intro}

Solutions of highly charged polymers are known to develop novel
structural properties due to the interplay between electrostatic
interactions and entropic effects \cite{gelbart}. For example, it
has been shown that for most charged biopolymers such as DNA,
filamentous (F-)actin, and various viruses \cite{DNA-collapse},
electrostatic correlations in the vicinity of the macroions caused
by multivalent counterions---ions of opposite charge---can lead to
like-charged attraction \cite{levin}. This attraction most often
destabilizes the polyelectrolyte solution and leads to the
formation of collapsed bundles.

In a recent experiment with F-actin, Wong {\em et al.} observed
that when the density of multivalent counterions is not yet
sufficient to trigger complete collapse, the uniform solution can
become unstable and phase separate into coexisting low- and
high-density phases \cite{gerard}. This experiment is perhaps a
most direct manifestation of the peculiar behavior of
polyelectrolytes in high density solutions: the dense phase is
characterized by multi-axial liquid crystalline behavior
\cite{itamar}, as well as exceedingly slow dynamics that leads to
the formation of an actin gel \cite{gerard}. A similar
self-assembly pattern has been observed in the structure of the
nuclear lamina---a charged filamentous integral membrane protein
network that provides a cytoskeletal support for the nuclear
membrane \cite{alb}.

Another interesting example is the observation of slow modes in
the dynamics of various high-density polyelectrolyte solutions in
the absence of salt, where it is suggested that correlations cause
the formation of starlike complexes with considerably slow
dynamics \cite{slow,muthu}. These examples raise the issue that
strong many-body Coulombic correlations for extended
(non-pointlike) charged objects could lead to novel features that
are poorly understood.

Based on such experimental evidence, one can broadly identify two
characteristic features in this class of problems: the novel
orientational ordering, and the slow dynamics. As a first step
towards understanding such properties, one should note the {\em
frustration} that is inherent structurally in the collection of
like-charged rodlike objects: As they come together, they reorient
themselves to minimize the total energy of the system. Here, the
long range nature of the interaction presents them with a
multitude of configurations that are good minimum energy
candidates and yet are far apart in the configuration space. This
picture is partly supported by the fact that there are so many of
them. This picture is supported by the fact that in the presence
of (a considerable amount of) salt the above effects are washed
out, as screening can eliminate the frustration by reducing the
effective range of the mutual repulsion.

\begin{figure}
\includegraphics[width=.8\columnwidth]{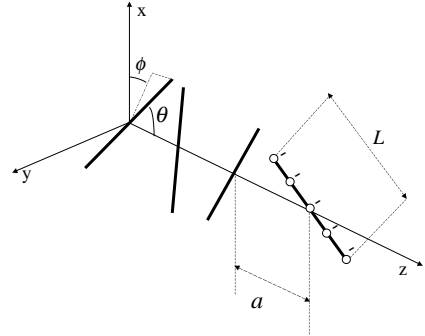}
\caption{The schematic configuration of charged rods. The centers
of the rods of equal length $L$ are fixed on a chain of spacing
$a$. Each rod has two rotational degrees of freedom, and a uniform
distribution of point charges.} \label{fig:schematic1}
\end{figure}

\begin{figure}
\includegraphics[width=.8\columnwidth]{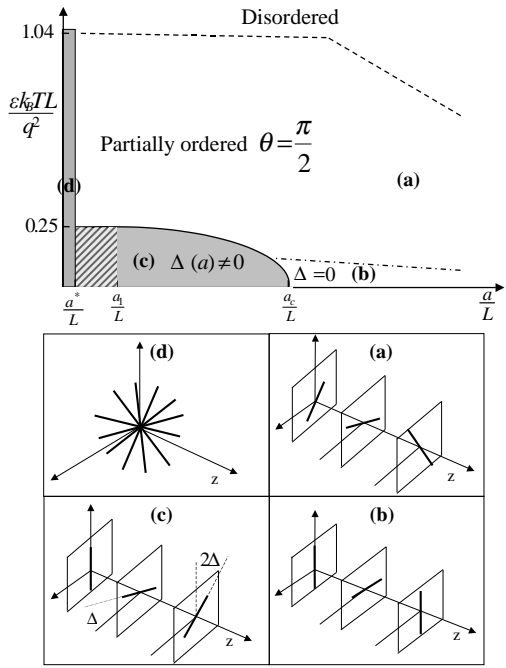}
\caption{Phase diagram of the system in the $a-T$ plane. Dashed
line: The crossover transition line from a disordered regime to
partially ordered one [region (a)]. Dash-dotted line: The
crossover transition line from a partially ordered regime to the
staggered ground-state of the low-density regime [region (b)].
Solid line: First order transition line from a partially ordered
regime to the chiral ordered phase of the intermediate-density
case [region (c)]. Hashed region: The domain in which a multitude
of transitions occur to phases with various kinds of orderings.
Region (d): The ``sea urchin'' phase. Schematic configuration of
rods corresponding to the different parts of the phase diagram is
indicated in the lower panel. The values of the various transition
points indicated in the phase diagram are $a_c=0.40\,L$,
$a_1=0.10\,L$, and $a^{*}\simeq 0.02\,L$.}\label{fig:phase1}
\end{figure}

Here, we set out to study the collective behavior of
polyelectrolytes at close separations using Monte Carlo (MC)
simulation, with the primary goal of understanding the complexity
that electrostatic frustration can bring about. We consider a most
simple model system of similarly charged rigid rods of equal
length $L$ with their centers forming a regular lattice of spacing
$a$ on a chain, as shown in Fig.~\ref{fig:schematic1}. We find a
variety of collective behaviors, and map out the phase diagram for
the system as a function of temperature and lattice spacing
(inverse density of the rods), as shown in Fig.~\ref{fig:phase1}.
In the low-density regime, the rods are found to order in a
staggered way upon decreasing the temperature, such that they are
perpendicular both to their neighbors and to the axis that
connects their centers, through a two-stage crossover transition.

For intermediate densities, we find that a low-temperature ordered
phase appears through a first-order phase transition, in which
neighboring rods have a twist angle in addition to the ninety
degrees of the low-density phase. In the high-density regime, a
hierarchy of different phases is observed depending on the lattice
spacing, where there is a multitude of twist angles, a periodic
structure with a basis, and out-of-plane arrangements of the rods.
We also study the equilibrium relaxation properties of the system
and find that the appropriate order parameter in the system has an
anomalous relaxation characterized by a stretched-exponential
behavior, which is a feature also seen in systems with
frustration.

The rest of the paper is organized as follows. In Sec.
\ref{sec:model}, we introduce the model and outline the simulation
technique used. Section \ref{sec:results} is devoted to the
description of various parts of the phase diagram, which is
followed by discussions on the relaxation dynamics of the system
in Sec. \ref{sec:dynamics}. Some discussions and concluding
remarks are presented in Sec. \ref{sec:conclusion}. Finally, a
variant of the model that corresponds to polyelectrolyte combs is
discussed in the Appendix.

\section{The Model}     \label{sec:model}

We use the Metropolis algorithm to simulate a system of $N_r$ rods
with their centers fixed on a chain, and only their rotational
degrees of freedom $(\theta,\phi)$ left to explore \cite{theta}.
We assume that $N_q$ (odd) point-charges of the same sign are
distributed symmetrically on each rod (Fig.~\ref{fig:schematic1}).
The screened Coulomb interaction energy of the system of rods can
be written as:
\begin{equation}
U=\frac{q^2}{\varepsilon}\sum_{m>n}^{N_r}\sum_{i,j=1}^{N_q}\frac{e^{-
\kappa |{{\bf r}_{mi}-{\bf r}_{nj}}|}}{|{{\bf r}_{mi}-{\bf
r}_{nj}}|}, \label{U-def}
\end{equation}
where $\varepsilon$ is the dielectric constant of the medium,
$q=Q/N_q$ where $Q$ is the overall charge of the rod, ${\bf
r}_{mi}({\bf r}_{nj})$ is the position of charge $i(j)$ on rod
$m(n)$, and $\kappa^{-1}$ is the Debye screening length. Note that
the energy expression is invariant under the local transformation
$\theta\to\pi-\theta$ and $\phi\to\phi+\pi$ for each rod. We use
free boundary conditions to avoid complexities arising from the
Ewald-summation of extended charge distributions \cite{Ewald}.

For each value of the lattice constant $a$, we start the system
from a random configuration of the rods at a high temperature
(where the system is in the disordered phase) and gradually cool
down toward lower temperatures. At a given temperature, we start
the calculation of various thermal averages after equilibration
and take the last configuration at each temperature as the initial
configuration for a slightly lower temperature. We have run a wide
variety of simulations for different values of $N_r$, $N_q$, and
$a$, and found that all the qualitative results are robust.

\section{The Phase Diagram}    \label{sec:results}

\begin{figure}
\includegraphics[width=.8\columnwidth]{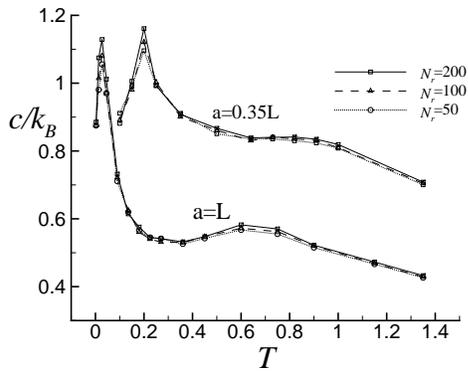}
\caption{The specific heat as a function of temperature $T$ [in
units of $q^2/(\varepsilon k_{\rm B} L)$] for (a) $a=L$ and (b)
$a=0.35\,L$, with three different sizes $N_r=50,\,100,\,200$ and
$N_q=3$.} \label{fig:specific}
\end{figure}

To capture the essence of the electrostatic frustration, we first
consider the salt-free case that corresponds to $\kappa=0$, and
comment on the effect of screening later. In
Fig.~\ref{fig:phase1}, we show the phase diagram of the system in
the plane of temperature and lattice constant for $N_{r}=100$ and
$N_{q}=3$.

In the low-density regime, where $a$ is greater than a critical
value $a_c=0.40\,L$, the rods appear to order on decreasing the
temperature through a two-stage crossover transition. The first
crossover (denoted by the dashed line in Fig.~\ref{fig:phase1})
takes the system to a partially ordered phase in which the angular
degree of freedom, $\theta$, freezes to $\frac{\pi}{2}$, and the
rods fluctuate freely only within the resulting parallel plates
[region (a) in Fig.~\ref{fig:phase1}]. The ordering in the $\phi$
degree of freedom occurs through another crossover at a lower
temperature (denoted by the dash-dotted line in
Fig.~\ref{fig:phase1}). The crossover transition temperatures are
determined from the specific heat of the system as a function of
temperature, which develops two rounded peaks. In Fig.
\ref{fig:specific}a, the specific heat of the system is plotted as
a function of the temperature $T$, which is made dimensionless by
the combination $q^2/(\varepsilon k_{\rm B} L)$, at $a=L$ for
three different sizes $N_r=50,\,100,\,200$ and $N_q=3$. Each data
point is the average over five independent runs starting from
different initial random configurations. For each temperature,
$1.5\times 10^6$ MCS are used with the first $7.5\times 10^5$ MCS
being for equilibration. As can be seen from Fig.
\ref{fig:specific}a, the height of the peaks in the specific heat
do not change appreciably for various values of the system size
$N_{r}=50,\,100,\,200$.

The ground-state configuration corresponding to this regime is
given as [see Fig.~\ref{fig:phase1}b]:
\begin{equation}
\theta_{m}=\frac{\pi}{2}~~{\rm and}~~ \phi_{m}=m\times
\frac{\pi}{2}, \label{regimebGSC}
\end{equation}
where $m$ indicates the position of the centers of rods on the $z$
axis $(z_{m}=ma)$. This is in agreement with the two-body minimum
energy configuration of the rods \cite{adrian}. A multipole
expansion of the electrostatic energy expression in Eq.
(\ref{U-def}), which is presumably a good approximation in the
limit of $a \gg L$, reveals that the system of rods effectively
behaves as a collection of quadrupoles in an external staggered
electric field; hence the perpendicular ordering described by Eq.
(\ref{regimebGSC}). This picture also agrees well with the lack of
genuine phase transitions in this regime, as it is expected from
quadrupoles in a symmetry breaking external field.

In the intermediate-density regime, corresponding to $a^{*}<a<a_c$
where $a^{*}\simeq 0.025\,L$, lowering the temperature for each
value of the lattice constant causes a crossover transition from
disordered to partially ordered regime, which is followed by a
phase transition (solid line in the phase diagram in
Fig.~\ref{fig:phase1}) to low-temperature phases having chiral
order. A plot for the specific heat in this regime, corresponding
to $a=0.35\,L$, is shown in Fig. \ref{fig:specific}b for system
sizes $N_r=50,\,100,\,200$ and $N_q=3$. The data points are the
averages of five independent runs starting from different initial
random configurations. For each temperature $2\times 10^6$ MCS are
used, with $10^6$ MCS being for equilibration. It can again be
seen that the specific heat is not sensitive to the size of the
system.

\begin{figure}
\includegraphics[width=.8\columnwidth]{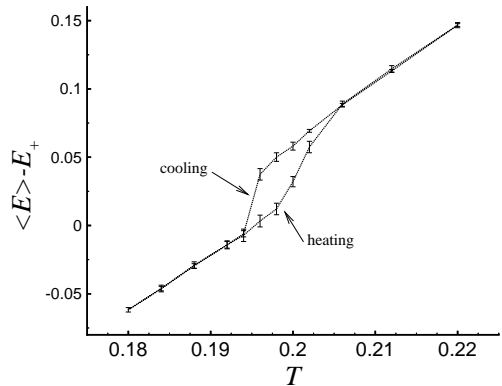}
\caption{The thermal average of energy per site [in units of
$q^2/(\varepsilon L)$] as a function of the temperature [in units
of $q^2/(\varepsilon k_{\rm B} L)$] upon successive cooling and
heating of the system, for $a=0.35\,L$, $N_r=100$, $N_q=3$. The
self-energy of the staggered ground-state configuration $E_+$ is
subtracted off from the energy. The hysteresis loop can be
observed, which indicates a first order transition from the
partially ordered regime to the chiral phase.}
\label{fig:hysteresis}
\end{figure}

\begin{figure}
\includegraphics[width=.8\columnwidth]{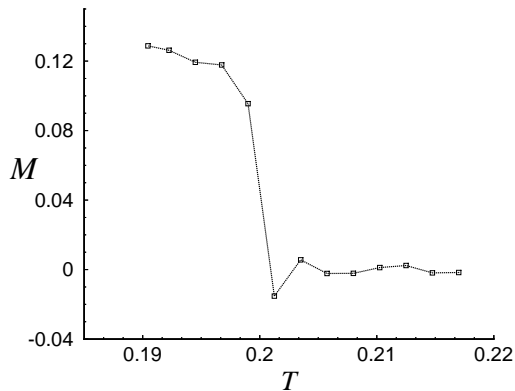}
\caption{The order parameter $M$ [defined in Eq. (\ref{M-def})]
versus temperature [in units of $q^2/(\varepsilon k_{\rm B} L)$]
for a system of $N_r=200$, $N_q=3$ at $a=0.35\,L$.}
\label{fig:delta}
\end{figure}

To understand the nature of the phase transition, we examined the
energy of the system upon successive cooling and heating. In Fig.
\ref{fig:hysteresis}, the thermal average of energy per site as a
function of the temperature [that is made dimensionless using
$q^2/(\varepsilon k_{\rm B} L)$] is plotted for $a=0.35\,L$
corresponding to successive cooling and heating of the system. In
this curve, $7.5\times 10^5$ MCS are used for equilibration and
$7.5\times 10^5$ MCS for ensemble averages, and it corresponds to
$N_r=100$, $N_q=3$. As can be seen from Fig. \ref{fig:hysteresis},
the average energy of the system shows hysteresis around the
transition temperature, which is a signature of a first order
phase transition. One can also define an order parameter that
could distinguish between the two phases and thus monitor the
phase transition as a function of temperature. A suitable choice
for the order parameter could be
\begin{equation}
M=\frac{1}{N_{r}-1}\sum_{n=1}^{N_{r}-1}\langle \sin\Delta_{n}
\rangle, \label{M-def}
\end{equation}
where $\Delta_{n}(t)=\phi_{n+1}(t)-\phi_{n} (t)-\frac{\pi}{2}$.
Figure \ref{fig:delta} shows a plot of this order parameter versus
temperature [in units of $q^2/(\varepsilon k_{\rm B} L)$] for a
system of $N_r=200$ and $N_q=3$ at $a=0.35\,L$. In this plot,
$10^6$ MCS are used for equilibration of the system and $10^6$ MCS
for thermal averages. It can be seen that the order parameter
develops a discontinuous jump across the transition temperature.
This is another indication that the transition across the solid
line in the phase diagram of Fig.~\ref{fig:phase1} is first order.
These observations, however, should be complemented with more
systematic investigations using energy histogram methods
\cite{LK,future}.

Below this transition, the system of rods appears to develop a
rich variety of orderings. For $a_1 < a < a_c$ where $a_1=0.1\,L$
[region (c) in the phase diagram in Fig.~\ref{fig:phase1}], the
array of rods orders in a chiral phase with the corresponding
ground-state configuration given as:
\begin{equation}
\theta_{m}=\frac{\pi}{2}~~{\rm and}~~ \phi_{m}=m\times
\left(\frac{\pi}{2}+\Delta_{g}(a)\right).\label{regimecGSC}
\end{equation}
Since the twist angle $\Delta$ is small in the vicinity of the
transition line, we can simplify our description by assuming that
$\Delta$ itself serves as the corresponding order parameter for
the phase transition from (a) and (b) to (c). While its value is
observed to undergo a finite jump just below the transition line,
the ground-state value $\Delta_{g}(a)$ is found to vanish
continuously at $a_c$, implying that the termination point of the
transition line at zero temperature corresponding to $a=a_c$ is a
critical point.

\begin{figure}
\includegraphics[width=.8\columnwidth]{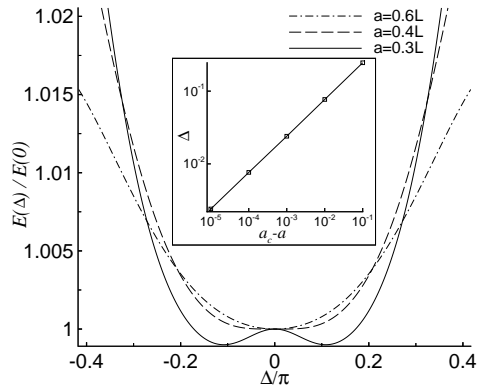}
\caption{A plot of the ground-state energy per site as a function
of $\Delta$ for $a=0.3\,L$, $a=a_c=0.4\,L$, and $a=0.6\,L$, with
$N_{r}=40$ and $N_{q}=3$. The value of $\Delta$ which minimizes
the energy of the system vanishes continuously at $a=a_c$. Inset:
The log-log plot of $\Delta_g$ versus $(a_{c}-a)$ gives a slope of
$\beta=0.500$. } \label{fig:energy}
\end{figure}

For a more direct study of the transition in the ground-state
configuration at $a_c$, we calculate the exact energy per site for
the system using the energy expression given in Eq. (\ref{U-def})
as a function of $\Delta$ for the configuration given in Eq.
(\ref{regimecGSC}). This means that we freeze the charged rods
into the configurations defined by Eq. (\ref{regimecGSC}) and
calculated in a point by point double summation the electrostatic
energy of the configuration for each value of $\Delta$. This
energy function, which is plotted in Fig.~\ref{fig:energy} for
different values of $a$, admits a Landau form and the
corresponding minimum energy solution $\Delta_{g}(a)$ vanishes at
$a_c$ as $\Delta_{g}=|a_{c}-a|^{\beta}$ with $\beta=0.500$ (see
inset of Fig.~\ref{fig:energy}). We have checked that the value of
$\beta$ does not depend on $N_q$ (even up to $N_q=1001$),
confirming the expectation from Landau theory that $\beta=1/2$.

In the intermediate-density regime when the rods are brought
closer to each other, different configurations that are close in
energy appear. This changes the behavior of the system at $a_1$
where the curve of energy versus $\Delta$ starts to have four
degenerate minima. For $a^*<a<a_1$ with $a^{*}\simeq 0.025\,L$
(hashed region in Fig.~\ref{fig:phase1}), while the rods still
respect the planar ordering, a multitude of transitions occur
taking the system through a variety of minimum energy
configurations. For example, upon increasing the density, a phase
appears in which successive cross-like structures formed by two
neighboring rods are rotated relative to each other around the
$z$-axis by an angle $\Delta^\prime$, which corresponds to a
chiral ordering with a basis. Finally, in the high-density region
where $a<a^*$, the rods are no longer confined in the planes.
While in this regime one also finds various kinds of orderings,
eventually the rods tend to approach a spherical configuration
that resembles a ``sea urchin'' [see Fig.~\ref{fig:phase1}(d)], as
$a$ approaches its smallest possible value $a_0$ that is set by
the thickness of the rods. Note that only a very small choice of
the cutoff length, such as our choice of $a_0= 10^{-6}\,L$, can
allow for these complicated high-density phases to exist.
Moreover, unlike the lower density cases, this part of the phase
diagram depends very sensitively on the choice of charge
distribution on the rods, namely the value of $N_q$ as well as
whether the charges are smeared or discrete.

\section{Relaxation Dynamics}    \label{sec:dynamics}

The low-temperature chiral ordering (in either of its various
forms) is not easily achieved in its complete form when the system
is cooled down from the disordered phase to temperatures below the
transition line. This means that the equilibrium dynamics of the
system involves a rather slow process, which we try to identify
and study here.

For $a_1<a<a_c$, the array of rods is decomposed into several
domains of the chiral structure with both positive and negative
values of $\Delta$, corresponding to the degenerate minima of
energy shown in Fig.~\ref{fig:energy}. The neighboring left-handed
and right-handed domains are regions that typically contain just a
few rods, and are linked via what can be thought of as {\em
kinks}. By definition, the kinks annihilate when they meet each
other or when they reach the boundaries of the system. We observe
that the density of the kinks is more or less fixed to about one
kink for every 40 rods or so, irrespective of the size of the
system.

These kinks are observed to have very slow dynamics, presumably
due to the electrostatic frustration, so that a considerably long
MC time is needed before all the kinks are annihilated and one of
the domains spans the entire system. The MC time needed for the
annihilation of the kinks increases with increasing the system
size. We also note that when the energy has four minima (for $a$
just below $a_1$) and two kinds of chiral structures are possible,
different kinds of kinks are observed depending on the type of the
two neighboring domains that are linked.
\begin{figure}
\includegraphics[width=.8\columnwidth]{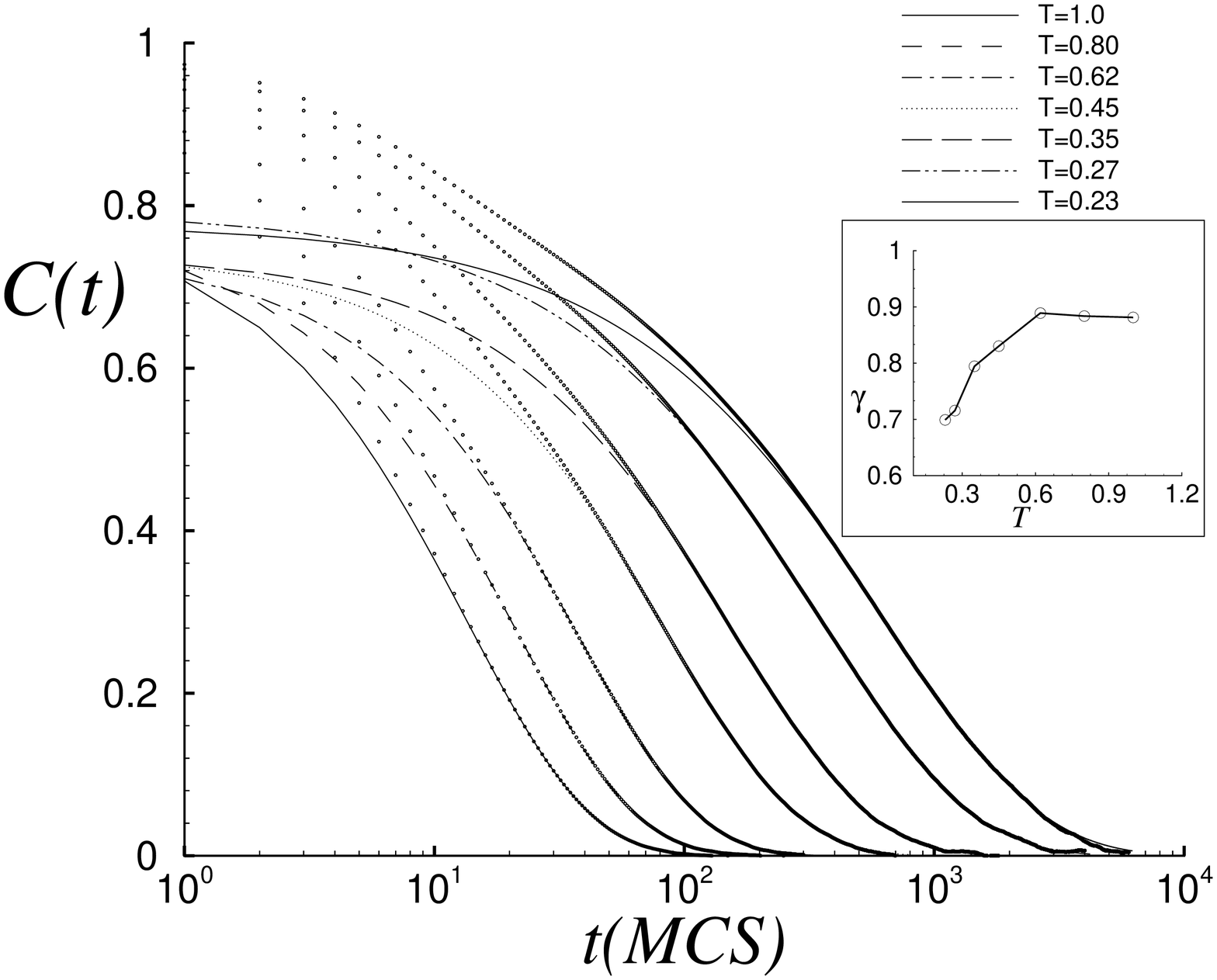}
\caption{Equilibrium relaxation function of the system for
$a=0.25\,L$ and various temperatures, reported in units of
$q^2/(\varepsilon k_{\rm B} L)$. The nonlinear fits to the
long-time behavior of the curves are of the stretched-exponential
form. Inset: A plot of $\gamma$ as a function of temperature.
}\label{fig:relax}
\end{figure}

To put the study of the slow dynamics in a more quantitative
framework, we probe the equilibrium relaxation properties of the
system by measuring the auto-correlation function for
$\Delta_{n}(t)$, which is defined as:
\begin{equation}
C(t)=\frac{\sum_{n=1}^{N_{r}-1}{\int dt'
\Delta_{n}(t+t')\Delta_{n}(t')}}{\sum_{n=1}^{N_{r}-1}{\int dt'
\Delta_{n}(t')\Delta_{n}(t')}}.\label{C(t)}
\end{equation}
To obtain $C(t)$ at each temperature $T$, we start the system from
a random configuration at high temperatures and quench it to $T$.
After running the system for a waiting time, $t_w$, we use the
system configuration $\phi_{n}(t)$ in all time steps from $t_w$ to
$t_{w}+t_{av}$ to calculate $C(t)$ using Eq. (\ref{C(t)}). For
long enough waiting times ($t_w \simeq 2\times10^6$ MC steps for
$N_{r}=100$)), the relaxation function $C(t)$ is independent of
$t_{w}$ as well as the system initial conditions, indicating that
there is no difference between averaging over time [Eq.
(\ref{C(t)})] and averaging over different initial conditions of
the system. Note that the waiting time is typically the time
needed for the kinks to annihilate.

In Fig.~\ref{fig:relax}, we show the system relaxation function
with logarithmic time scale for different values of temperature in
the case of $a=0.25\,L$ with $N_{r}=100$ and $N_{q}=3$. Using
nonlinear fits with three free parameters, we find that the
long-time part of the relaxation function is well represented by
the stretched-exponential form, $C_{0} \,\exp[-(t/\tau)^\gamma]$.
The exponent $\gamma$ saturates to a constant of the order of
$0.9$ in the high temperature regime, while it drops at lower
temperatures where the relaxation is exceedingly slower. The
stretched-exponential form for the relaxation is a feature that is
often encountered in the dynamics of frustrated systems
\cite{frustrated}.

\section{Discussion and Conclusion}    \label{sec:conclusion}

In a most simple realization of the system, we have shown that
charged linear objects in close separation can develop novel
collective orientational correlations as well as an unusually slow
dynamics, due to the frustration in the orientational degrees of
freedom. While in our simple model system we assume that the rods
are positioned in a lattice, we expect that removing this
constraint will cause the orientational correlations to manifest
themselves as liquid crystalline structures such as those observed
in Ref. \cite{gerard}. Geometrical frustration induced by the
dimensionality of the arrangement could also affect the
orientational ordering, and it remains to be seen what kind of
arrangement is favored by the liquid system itself, i.e., what is
the coupling between the positional- and the orientational
correlations in this frustrated system.

Let us examine the effect of screening by added-salt in the above
results. For a low concentration of added salt, the screening is
weak and we expect that the results reported here will hold true
so long as the screening length is the largest length in the
system. As the salt concentration is increased, however, the Debye
length begins to compete with the other length scales and this
could change the phase behavior of the system.

\begin{figure}
\includegraphics[width=.8\columnwidth]{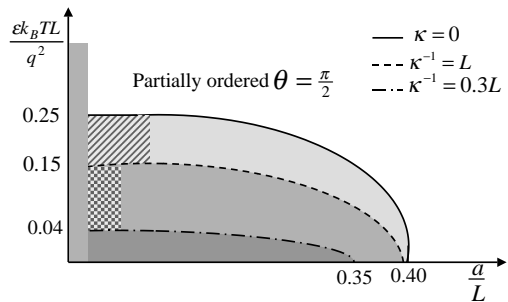}
\caption{First order transition line of the phase diagram for
different values of the screening parameter.} \label{fig:screen}
\end{figure}

We have performed simulations with the screened Coulomb
interaction of Eq. (\ref{U-def}) using different values for the
Debye screening length. We find that the general structure of the
phase diagram of Fig. \ref{fig:phase1} persists, while the size of
the chiral domain in the phase diagram shrinks as $\kappa$ is
increased. In Fig. \ref{fig:screen}, the first order phase
transition line is shown for three values of the Debye length for
$N_r=100$ and $N_q=3$. It can be seen that even for
$\kappa^{-1}=L$ the value of $a_c$ barely changes as compared to
its $\kappa=0$ value of $0.4\,L$. Our results suggest that
$\kappa^{-1} \sim a_c$ sets the threshold where screening can wipe
out the chiral ordering altogether.

We have neglected the effect of counterions in the above study.
This can be justified by noting that the polyelectrolytes are very
short rods and unless we choose extremely high values for $q$, it
is unlikely that they condense on the rods. This means that they
will move relatively freely in the vicinity of the linear
polyelectrolyte array, and merely participate in the Debye
screening of the interaction. This effect can be taken into
account by considering a local effective value for the Debye
length that incorporates the density of the counterions as well as
that of salt. Moreover, the presence of the counterions in an
effective ``cell'' around the array provides the necessary
neutralization of the system. In our treatment, however, we did
not need to care about the neutralization criterion because the
centers of the charged rods are fixed in space and in effect it is
only multipoles (not monopoles) that are interacting.

Finally, we note that the results obtained here are related to the
studies of the equilibrium configuration of ions in a confined
plasma, where even similar chiral phases are observed as the
confinement potential becomes anisotropic \cite{plasma}.

\acknowledgments

We are grateful to M. Deserno, R.R. Netz, D. Stauffer, T.A. Waigh,
and G.C.L. Wong for interesting discussions and comments.

\appendix

\section{End-grafted Charged Rods}   \label{sec:brush}

In this Appendix, we consider the case where the charged rods are
actually grafted at their ends instead of the mid-points.
This will make a model for a
polyelectrolyte comb, i.e. a backbone with a linear array of
charged side chains, which has in fact been synthesized and
studied \cite{comb,waigh}.

\begin{figure}
\includegraphics[width=.8\columnwidth]{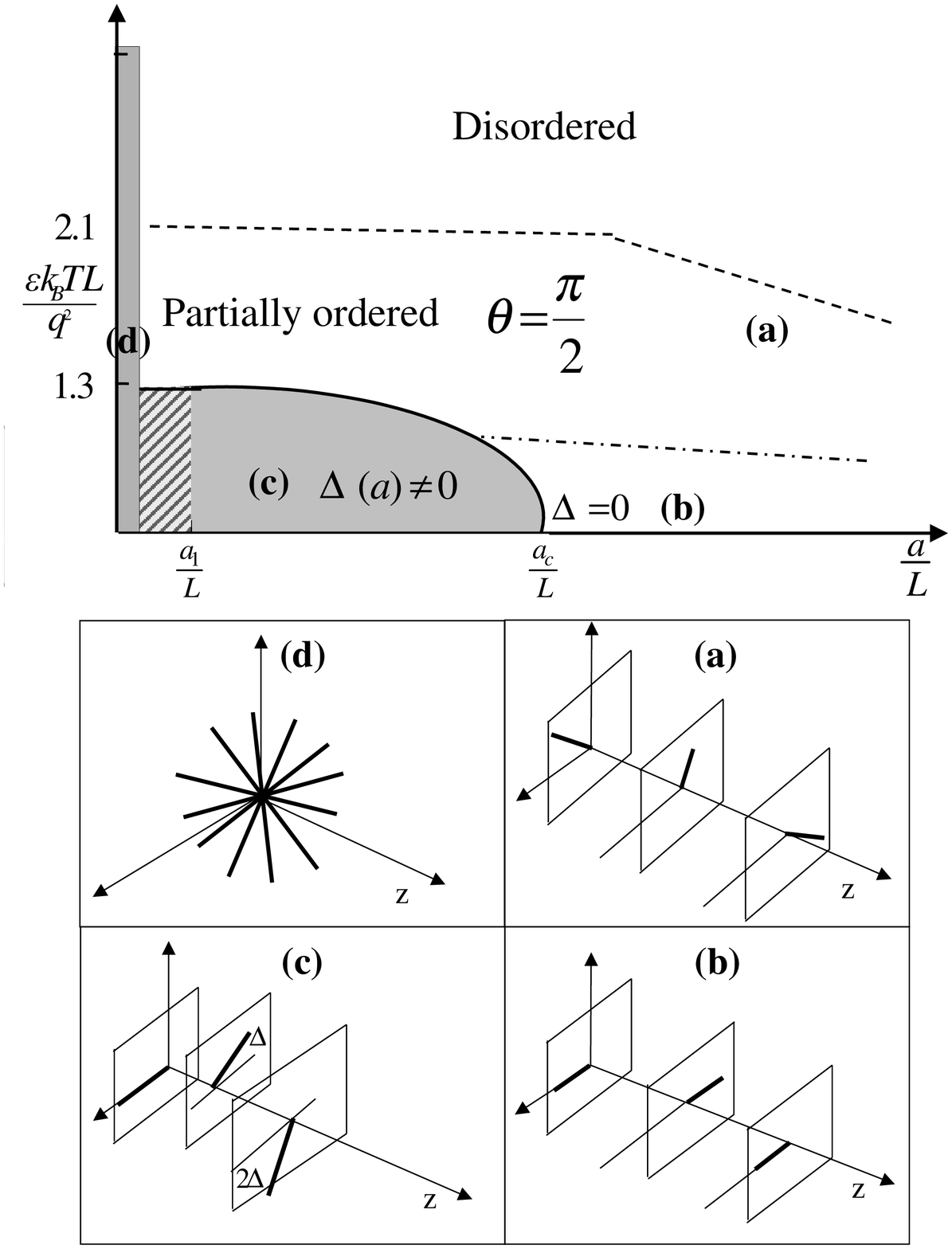}
\caption{Phase diagram of the polyelectrolyte comb model in the
$a-T$ plane. Dashed line: The crossover transition line from a
disordered regime to partially ordered one [region (a)].
Dash-dotted line: The crossover transition line from a partially
ordered regime to the staggered ground-state of the low-density
regime [region (b)]. Solid line: First order transition line from
a partially ordered regime to the chiral ordered phase of the
intermediate-density case [region (c)]. Hashed region: The domain
in which a multitude of transitions occur to phases with various
kinds of orderings. Region (d): The ``sea urchin'' phase.
Schematic configuration of rods corresponding to the different
parts of the phase diagram is indicated in the lower panel. The
values of the various transition points indicated in the phase
diagram are $a_c=1.40\,L$ and $a_1=0.3\,L$.}\label{fig:phase2}
\end{figure}

We have performed simulations on such a system and have found the
same general behavior as the previous case. The corresponding
phase diagram for the case of $N_q=7$ is shown in Fig.
\ref{fig:phase2}, where the transition points are found to be
$a_c=1.40\,L$, and $a_1=0.3\,L$. Note that the region (b) in Fig.
\ref{fig:phase2} corresponds to rods that are anti-parallel to
their neighbors, which is a direct consequence of the
non-vanishing dipole moment of rods that are grafted at one end as
opposed to those that are grafted from the mid-point.
Consequently, region (b) in this case corresponds to dipoles in a
staggered field. Similarly, in the chiral phase the neighboring
rods are out of phase by $\pi$ (instead of $\pi/2$) plus a
residual twist angle. We also note that the transitions are
occurring at larger values of the reduced temperature and lattice
spacing, because of the stronger dipolar interactions.

\begin{thebibliography}{99}
\bibitem{theta}
In sampling the unit sphere described by $(\theta,\phi)$
uniformly, one should be careful to include the $\sin \theta$
factor in the measure. In most interesting domains in the phase
diagram of Fig.~\ref{fig:phase1}, however, the tendency to keep
the rods in-plane makes this factor dispensible, as our explicit
checks have shown.
\end{thebibliography}
\end{document}